\documentclass[11pt,preprint]{aastex}
\usepackage{times}
\let\oldfootsep=\footnotesep
\newcommand\ltsima{$\; \buildrel <\over\sim \;$}
\newcommand\simlt{\lower.5ex\hbox{\ltsima}}
\newcommand\gtsima{$\; \buildrel >\over\sim \;$}
\newcommand\simgt{\lower.5ex\hbox{\gtsima}}

\newcommand\msun { \rm {M_\odot}}

\newcommand\vperp{v_{\rm \perp}}
\newcommand\vperpbold{\bf v_{\rm \perp}}
\newcommand\vp {\tilde{v}}
\newcommand\vpbold{\tilde{\bf v}}

\newcommand\pac{Paczy{\'n}ski }

\newcommand\eclong{\lambda}
\newcommand\eclat{\beta}  %% This is conventional
\newcommand\vs {{\bf v}_S}
\newcommand\vl {{\bf v}_L}

\newcommand\that {\widehat{t}}

\newcommand\kms {\,{\rm km/s} }

\shorttitle{MACHO-99-BLG-22}
\shortauthors{Bennett et al.}

\begin{document}


\title{The Microlensing Event MACHO-99-BLG-22/OGLE-1999-BUL-32: \\
       An Intermediate Mass Black Hole, or a Lens in the Bulge}


\author{
    D.P.~Bennett\altaffilmark{1},
    A.C.~Becker\altaffilmark{2},
    J.J.~Calitz\altaffilmark{3},
    B.R.~Johnson\altaffilmark{4},
    C.~Laws\altaffilmark{5},
    % D.~Minniti\altaffilmark{14},
    J.L.~Quinn\altaffilmark{1},
    S.H.~Rhie\altaffilmark{1}, and
    W.~Sutherland\altaffilmark{6}
        }

\clearpage


\altaffiltext{1}{Department of Physics, University of Notre Dame, IN 46556}
\altaffiltext{2}{Bell Laboratories, Lucent Technologies, 600 Mountain Avenue,
        Murray Hill, NJ 07974}
\altaffiltext{3}{Physics Department,
    University of the Free State, Bloemfontein 9300, South Africa}
\altaffiltext{4}{Tate Laboratory of Physics, University of Minnesota,
  Minneapolis, MN 55455}
\altaffiltext{5}{Departments of Astronomy and Physics,
    University of Washington, Seattle, WA 98195}
\altaffiltext{6}{Royal Observatory, Blackford Hill, Edinburgh EH9 3HJ}
% \altaffiltext{14}{Depto. de Astronomia, P. Universidad Catolica, Casilla 104,
%         Santiago 22, Chile}


\clearpage

\begin{abstract}
We present an re-analysis of the longest timescale gravitational
microlensing event discovered to date: MACHO-99-BLG-22/OGLE-1999-BUL-32,
which was discovered by both the MACHO and OGLE microlensing alert systems.
Our analysis of this microlensing parallax event includes a likelihood
analysis of the lens position based upon a standard model of the Galactic
velocity distribution, and this implies that the lens could be a black hole
of $\sim 100\msun$ at a distance of a few hundred parsecs in the Galactic disk
or a massive stellar remnant (black hole or neutron star) in the Galactic 
bulge. Our new analysis includes data from the MACHO, GMAN, and MPS
collaborations in addition to the OGLE data used in a previous analysis
by Mao et al (2002). The crucial feature that distinguishes our
analysis from that of Mao et al is an accurate constraint on the 
direction of lens motion and an analysis of the implications
of this direction.
\end{abstract}


%\keywords{}


\section{Introduction}
\label{intro}

The abundance of long timescale microlensing events towards the
Galactic bulge has remained a puzzle \citep{hangould-mspec,bennett-parbh}
since shortly after the first systematic analyses of Galactic bulge 
microlensing events were reported \citep{ogle-tau,macho-bulge45}. 
A long time scale event can be caused by a massive lens, a slow relative
transverse velocity between the lens and source, or by large source star-lens
and lens-observer separations. Thus, the excess of
long timescale events could be caused by errors in our assumptions
regarding the phase space distribution of the source stars or the lenses,
or they could be caused by an unexpectedly large population of massive
lenses. However, the phase space distribution of stellar mass objects
in the Galactic disk and bulge is tightly constrained by observations,
whereas little is known about the mass function of massive stellar remnants.
Thus, it is possible that the excess of long timescale events is due to 
a population of black holes and neutron stars that is larger than expected.

This possibility has recently received observational support from the
MACHO \citep{bennett-parbh} and OGLE \citep{ogle-99b22} collaborations
which have identified individual candidate black hole microlensing
events based upon the detection of the microlensing parallax effect
\citep{refsdal-par,gould-par1,macho-par1}. This refers to the
detection of light curve features due to the Earth's accelerating motion
around the Sun, and it is often detectable in long timescale microlensing
events. To date, some 12 clear cases of this effect have been reported
by the MACHO \citep{macho-par1,bennett-parbh,becker-thesis}, OGLE 
\citep{mao-par,ogle2000bul43,ogle-par2,ogle-99b22}, MOA \citep{moa-par}
and PLANET \citep{planet-er2000b5} collaborations. These microlensing
parallax events are useful because they allow one to learn more about
the nature of the lens than can be learned for most microlensing events.
For most microlensing events, the only measurable parameter which constrains
interesting properties of the lens is the 
Einstein diameter crossing time, $\that$,  which depends on the lens
mass ($M$), distance ($D_\ell$), and the magnitude of the
transverse velocity ($\vperp = |\vperpbold |$). It
is given by
\begin{equation}
   \that = {2 R_E\over \vperp} = {4\over \vperp c}
           \sqrt{GM D_\ell (D_s - D_\ell )\over D_s} \ ,
   \label{eq-that}
\end{equation}
where $D_s$ refers to the distance to the source (typically $\sim 8\,$kpc for a
bulge source), and $R_E$ is the radius of the Einstein Ring. In a microlensing
parallax event, it is also possible to measure $\vp$, which is
the lens star's transverse speed projected to the Solar position, given by
\begin{equation}
\vpbold = \vperpbold D_s/(D_s-D_\ell) \ . \label{eq-vp}
\end{equation}
The measurement of both $\that$ and $\vp$ give us two measurements for
three unknowns ($M$, $\vperp$, and $D_\ell$), which allows us to solve 
for the mass as a function of distance:
\begin{equation}
   M = {\vp^2 \that^2 c^2 \over 16 G} {D_s-D_\ell \over D_\ell D_s}
     = {\vp^2 \that^2 c^2 \over 16 G} {1-x \over x D_s} \ ,
   \label{eq-m}
\end{equation}
where $x=D_\ell/D_s$. For the MACHO-99-BLG-22/OGLE-1999-BUL-32 microlensing
event, \citep{ogle-99b22} have used Eq.~\ref{eq-m} to argue that the
lens mass is well above the measured mass of neutron stars ($\sim 1.4\msun$)
for plausible values of $x$.

\citet{bennett-parbh} have also made use of the measured direction of the
projected velocity, $\vpbold$, in a likelihood analysis employing the
velocity and density distributions of a standard Galactic model. This
allows a likelihood estimate for the lens distance, $x$, which can be 
converted to an estimate of the lens mass using Eq.~\ref{eq-m}. This
analysis has revealed two other candidate black hole lenses, MACHO-96-BLG-5
and MACHO-98-BLG-6, as well as several other events with possible neutron
star lenses. In this paper, we add additional data to the analysis of the
MACHO-99-BLG-22/OGLE-1999-BUL-32 microlensing event, and show that this 
results in tighter constraints on the microlensing parallax parameters. We 
then apply the likelihood analysis developed in \citet{macho-par1} and
\citet{bennett-parbh} to estimate the likely distance and
mass of the lens. 

\section{The Data Set and Previous Analysis}
\label{sec-data}

% peak: 2741.9   A = 26.8594   July 5, 1999 21:36 GMT

The MACHO-99-BLG-22/OGLE-1999-BUL-32 microlensing event was discovered
and announced by the MACHO Alert system\footnote[1]{A catalog of
MACHO alert events is available at
{\tt http://darkstar.astro.washington.edu/} along with some information
regarding each event. However, the reported timescales of these events
are systematically underestimated because they come from fits that do
not allow for blending.} \citep{macho-alert},
and it was independently discovered by the
OGLE early warning system \citep{ogle-ews} about two months later.
Following the MACHO alert, this event was added to the observing schedule
for the GMAN \citep{becker-thesis} and MPS 
\citep{mps-98smc1} microlensing follow-up groups. MACHO and GMAN observations
ended in 1999, while MPS and OGLE observations continued through 2000.
Photometry from the images of this event is particularly challenging
because of severe crowding within $\sim 1.5^{\prime\prime}$ of the
source star. This makes it difficult for point spread function fitting,
crowded field photometry programs such as DOPHOT \citep{dophot},
DAOPHOT/ALLFRAME \citep{allframe}, or SoDOPHOT \citep{macho-sod,macho-calib}
to obtain accurate photometry for this event. For the GMAN and MPS
images, we were able to solve this difficulty by carefully selecting
template images with sub-arc second seeing, but no such images are
available for the MACHO data. There is a great deal of scatter
in the MACHO-red data which appears to be inconsistent with the other data
sets. A similar problem was seen in the MPS data, but it disappeared
when the MPS data was reduced with a better seeing image as a template.
This suggests that the photometric problem with the MACHO-red data is
due to close neighbor stars that are not quite
resolved in the MACHO-red template. The MACHO-blue data has much less scatter, 
and so we have decided to use only the MACHO-blue data in our analysis.

The data set presented in this paper consists of 609 MACHO-Blue band 
measurements\rlap,\footnote[2]{The MACHO Project survey data are
available from {\tt http://wwwmacho.mcmaster.ca/} and 
{\tt http://wwwmacho.anu.edu.au}.}
245 OGLE I-band measurements from their catalog of difference imaging
measurements \citep{ogle-dia-ev}, 179 MPS R-band observations
from the Mt.~Stromlo 1.9m telescope, and 21 GMAN R-band observations from
the CTIO 0.9m telescope. The MPS data are reduced using a slightly
modified version of the MACHO photometry code, SoDOPHOT. The
photometric errors reported by SoDOPHOT are modified by adding
a 1.0\% assumed systematic uncertainty in quadrature. The
CTIO data were reduced with the ALLFRAME package \citep{allframe},
with the error estimates multiplied by a factor of 1.5 to account for
systematic errors.

A microlensing parallax fit for this event was previously published
by \citet{ogle-99b22}. However, \citet{ogle-99b22} present their
fit results in a coordinate system that is inconvenient for
Galactic bulge microlensing events. The difficulty is that all
geometrical quantities are projected into the ecliptic plane, rather
than a plane perpendicular to the line of site to the lens that is
used for most discussions of microlensing events. Since the ecliptic
plane intersects the Galactic bulge, this projection can be very nearly
singular for Galactic bulge events. Instead, if we project the Earth's
motion in the ecliptic plane to the plane perpendicular to the line of
site (as in \citet{macho-par1}, for example), then there is no singularity.
Event MACHO-99-BLG-22/OGLE-1999-BUL-32
is about 5 degrees from the ecliptic plane
($\eclong = 271.122^\circ$, $\eclat = -5.142^\circ$). This means that
the transformations of the error bars from the coordinate system
of \citet{ogle-99b22} to our coordinate system are highly non-linear,
and it means that the $1\,\sigma$ error bars reported in \citet{ogle-99b22}
cannot be used to estimate the $2\,\sigma$ error bars. Because of this
difficulty and the fact that \citet{ogle-99b22} use a data set which differs
from the publicly available one, we will not attempt a detailed comparison
of our fit parameters with their results.

\section{Microlensing Parallax Analysis}
\label{sec-anal}

The magnification for a normal microlensing event with no detectable 
microlensing parallax is given by
\begin{equation}
  A(t) = {u^2 + 2 \over u\sqrt{u^2+4}} \ ;
\ \  u(t) \equiv \sqrt{u_0^2 + [2(t-t_0)/\that )]^2 } \ ,
   \label{eq-Astd}
\end{equation}
where $t_0$ is the time of closest approach between the angular positions
of the source and lens, and $u_0 = b/R_E$ where $b$ is the distance of the
closest approach of the lens to the observer-source line. Eq. (\ref{eq-Astd})
can be generalized to the microlensing parallax case
\citep{macho-par1} by assuming the perspective of an observer located at
the Sun. We can then replace the expression for $u(t)$ with
\begin{eqnarray}
       u^2(t) & = & u_0^2 + \omega^2 (t-t_0)^2
        + \alpha^2 \sin^2[\Omega(t-t_c)] \nonumber \\
& & + 2\alpha \sin[\Omega(t-t_c)]\,[\omega (t-t_0) \sin\theta
+ u_0\cos\theta ] \nonumber \\
& & + \alpha^2 \sin^2\eclat\,\cos^2[\Omega(t-t_c)]
+ 2\alpha \sin \eclat\,\cos[\Omega(t-t_c)] \,
        [\omega (t-t_0) \cos\theta - u_0\sin\theta]
   \label{eq-upar}
\end{eqnarray}
where $\eclong$ and $\eclat$ are the ecliptic longitude and latitude,
respectively,
$\theta$ is the angle between $\vperp$ and the North ecliptic axis,
$\omega = 2/\that$, and $t_c$ is the time when the Earth is closest
to the Sun-source line.
The parameters $\alpha$ and $\Omega$ are given by
\begin{equation}
\alpha = { \omega (1 {\rm AU}) \over \vp }
                \left( 1 - \epsilon \cos [\Omega_0 (t-t_p)]
\right) \ , \label{eq-al}
\end{equation}
and
\begin{equation}
\Omega (t-t_c) = \Omega_0 (t-t_c) + 2 \epsilon \sin[\Omega_0 (t-t_p)] \ ,
   \label{eq-Omega}
\end{equation}
where $t_p$ is the time of perihelion, 
$\Omega_0 = 2 \pi \,{\rm yr}^{-1}$, $\epsilon = 0.017$ is the
Earth's orbital eccentricity.

\begin{figure}
\plotone{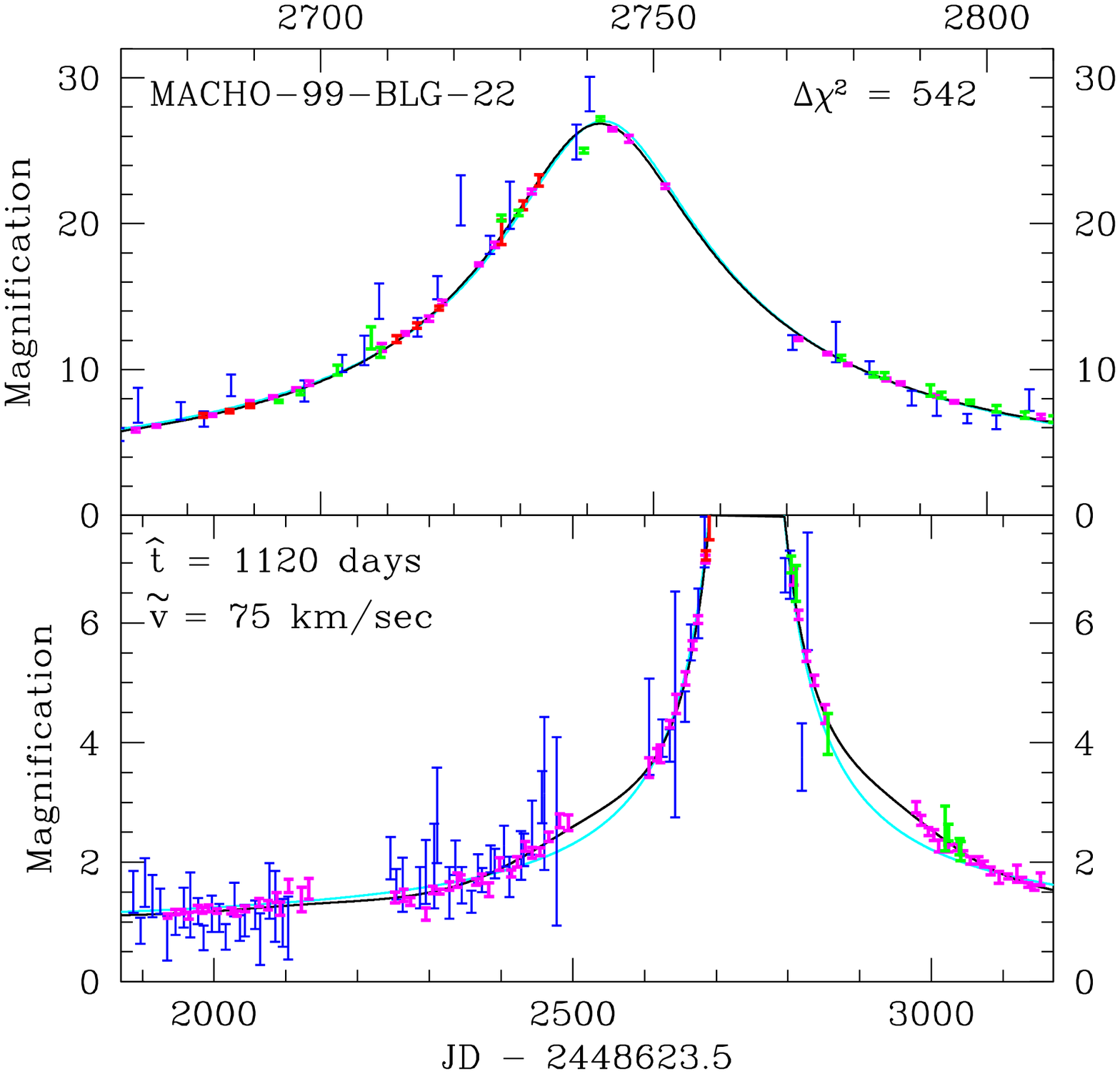}
\caption{
The light curve of the MACHO-99-BLG-22/OGLE-1999-BUL-32 microlensing
event with data from MACHO, OGLE, MPS and GMAN. The blue, green, red,
and magenta data points indicate data from the MACHO-Blue band, the
Mt. Stromlo 1.9m R-band,the CTIO 0.9m R-band, and the OGLE I-band,
respectively. The solid curve shows the best fit microlensing parallax
event while the cyan curve shows the best fit standard microlensing
light curve. In the upper panel, the data have been averaged into
4-day bins, while in the lower panel the data have been averaged into
10-day bins.
\label{fig-lc}}
\end{figure}

The microlensing parallax fits were performed with the MINUIT routine
from the CERN Library, and the results are summarized in 
Table~\ref{tbl-fitpar}, and are shown in Fig.~\ref{fig-lc}
The first four fit parameters listed in Table~\ref{tbl-fitpar} are
fit blend fractions for each of the four data sets. Because the OGLE
data was reduced with a difference imaging method, there is no photometric
reference flux value to use to define $f_{\rm OGLE} = 1$. Therefore,
we select the fit flux for the best parallax fit to define $f_{\rm OGLE} = 1$.

Although the comparison with the parallax fit of \citep{ogle-99b22} is
complicated by their peculiar choice of coordinates, we can compare to
their $\that$ and $\vp$ values since these are not affected by their
coordinate system. They find $\that = 1280 {+140\atop -110}\,$days and
$\vp = 79\pm 16\,$km/s, while we find $\that = 1120\pm 90\,$days
and $\vp = 75 {+7\atop -9}\,$km/s. We find that the parallax fit gives
an overall $\chi^2$ improvement of 541.7 which compares to the $\chi^2$ 
improvement of 298.1 found by \citep{ogle-99b22}. Thus, our results seem to be
somewhat more accurate, but consistent with the Mao et al fit.

The microlensing parallax fit parameters can be used to determine
the lens mass as a function of distance following Eq.~\ref{eq-m} as shown
in Fig.~\ref{fig-masslike}.

\subsection{Likelihood Distance and Mass Estimates}
\label{sec-likeli}

We can make use of our knowledge of the phase space distribution of
lens objects in the Galactic disk and bulge to further constrain
the distance and mass of the lens because the probability to
obtain the measured value of $\vpbold$ is quite sensitive to the
distance to the lens. Following \citet{macho-par1} and
\citet{bennett-parbh}, we can define a likelihood function
\begin{equation}
 L(x;\vpbold) \propto \sqrt{x (1-x)} \, \rho_L (x) \, \vp (1-x)^3 
\int f_S(\vs) \, f_L( (1-x) ({\bf v}_\odot + \vpbold) + x \vs ) \, d\vs , 
\end{equation}
where $\rho_L$ is the density of lenses at distance $x = D_\ell/D_s$,
and the integral is over combinations of source and lens velocities
giving the observed $\vpbold$. $\vs$ and
$\vl=(1-x) ({\bf v}_\odot + \vpbold) + x \vs$ are 
the 2-D source and lens velocity distribution functions.
We assume the following Galactic parameters:
a disk velocity dispersion of $30 \kms$ in each direction, 
a flat disk rotation curve of $200 \kms$, and a bulge 
velocity dispersion of $80 \kms$ with no bulge rotation. 
The density profiles are a standard double-exponential disk and a 
\citet{hangould-mspec} barred bulge, and the source star is assumed to
reside the the Galactic bulge at a distance of $8.5\,$kpc which is consistent
with the measured radial velocity, $v_r = +38\pm 2\,$km/sec of the source
star \citep{cook-spectra}.

\begin{figure}
\plotone{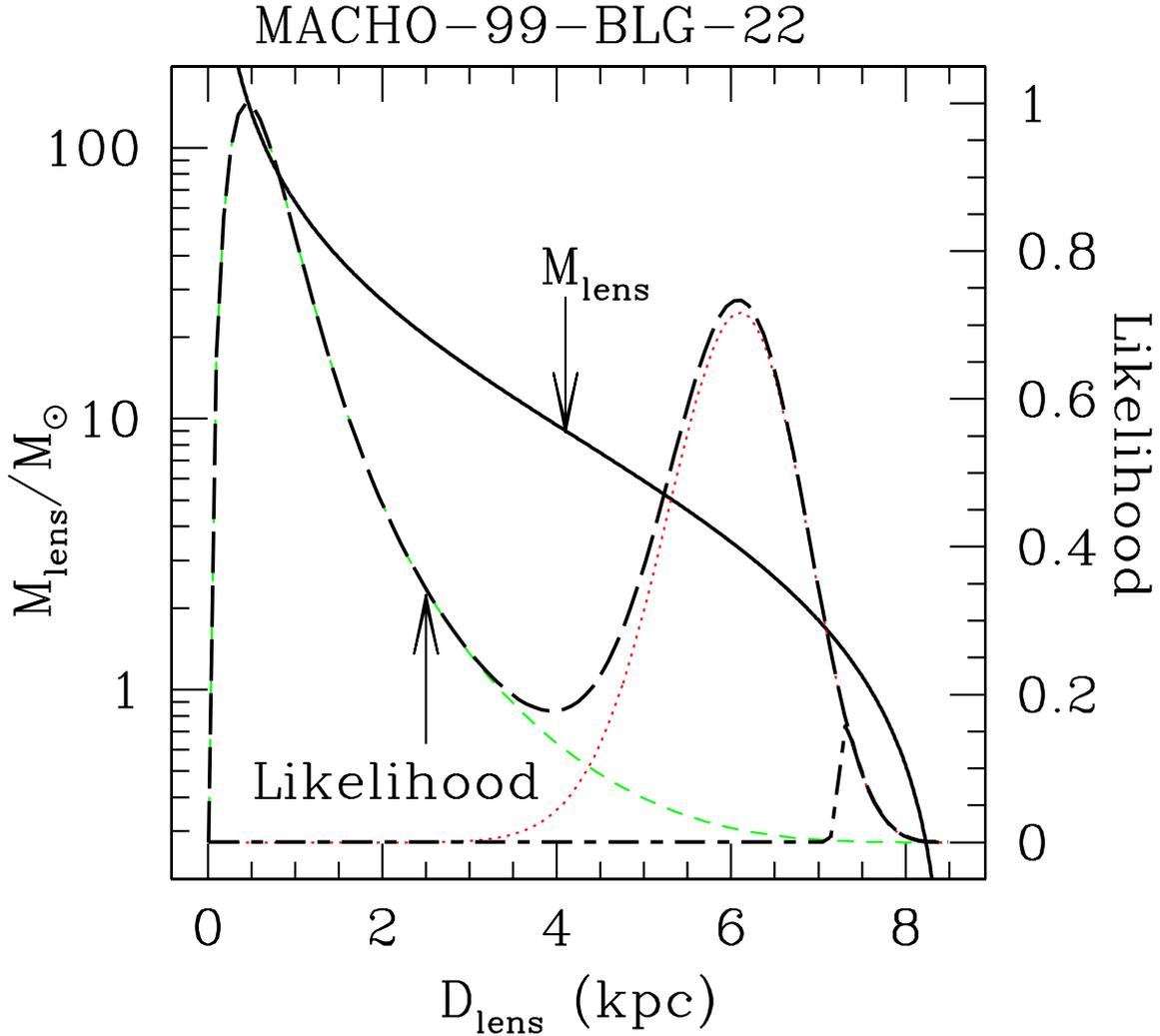}
\caption{
The mass vs.~distance relation (solid curves) for the
MACHO-99-BLG-22/OGLE-1999-BUL-32 microlensing event is
shown along with the likelihood functions (long dashed curves)
computed assuming a standard model for the Galactic phase space distribution.
The (green) short dashed curve is the contribution from disk lenses, and
the (red) dotted curve is the contribution from bulge lenses. The
alternating long and short dashed curve is the 
likelihood function which includes the constraint
on the brightness of a main sequence lens star. A main sequence lens
is quite unlikely for this event.
\label{fig-masslike}}
\end{figure}

The resulting likelihood functions for 
$D_{\ell}$ is shown as the long-dashed curves in 
Fig.~\ref{fig-masslike}. This differs from the likelihood functions for the
six events presented by \citet{bennett-parbh} because it has two peaks.
This is because the measured direction of $\vpbold$ is nearly opposite of the
direction of Galactic disk rotation. For a typical Galactic disk lens, we
expect $\vpbold$ to be roughly parallel to the disk rotation direction as
is observed for the six MACHO events analyzed in \citet{bennett-parbh}.
Since the bulge is assumed to have little or no rotation, the average
line-of-site to a bulge lens is stationary at the position of the source.
(A rotating bar-shaped bulge could imply an average bulk motion at
along some lines-of-sight.)
The flat rotation curve of the Galactic disk will cause stars orbiting 
interior to the Solar circle to have a higher angular velocity than the
Sun, and so the preferred projected velocity, $\vpbold$ will be in the
direction of Galactic disk rotation. However, the probability for $\vpbold$
to be in the anti-preferred direction is not small \citep{bennett-parbh},
so the fit parameters for MACHO-99-BLG-22/OGLE-1999-BUL-32 should not
be considered unlikely.

The double peaked nature of the likelihood function shown in 
Fig.~\ref{fig-masslike} can easily be understood if we consider the ways
in which a $\vpbold$ parallel to the Galactic disk rotation can be avoided.
One possibility is for a disk lens to be very close to us. In this case, the
average Galactic angular velocity at the lens position is only slightly larger
than that of the Sun, and so the velocity dispersion of the disk will
give a substantial population of potential lens objects with
retrograde $\vpbold$ values. The other possibility is that the lens
resides in the Galactic bulge where there is little or no systematic
rotation. In this case, with both the lens and source in the bulge, the
Galactic orbital motion of the Sun gives a preference for a 
retrograde $\vpbold$ value, but the preference is fairly small because
the lens must be much closer to the source than to the Sun.
The two likelihood function peaks can be then understood as being due to
disk and bulge lenses. This is clear from the decomposition of the
likelihood function shown in Fig.~\ref{fig-masslike}. For a disk lens, the
likelihood function peaks at a mass of $130\msun$, which is about an
order of magnitude above the most massive of the known stellar mass 
black holes in x-ray binary systems. If the lens is in the bulge, however,
the most likely mass is only about $4\msun$.

One common way to interpret likelihood functions is the Bayesian method,
in which the lens mass (or distance) probability distribution is given
by the likelihood function times a prior distribution, which represents
our prior knowledge of the probability distribution. In our case, the
likelihood function represents all of our knowledge about the lens mass
and location, so we select a uniform prior. With a uniform prior, the
likelihood function becomes the probability distribution, and
we are able to 
calculate the lens mass confidence levels listed in Table~\ref{tbl-mass}.
Also included in this table are the lens mass confidence levels if the
lens is known to be a disk or a bulge star. If the lens resides in the
disk, Table~\ref{tbl-mass} indicates that the lens mass should be in
the range $16\msun < M < 172\msun $ at $1\,\sigma$ and
$6.3 < M < 900\msun $ at $2\,\sigma$. This wide range is due to 
the fact that the distance to a disk lens is not well constrained. If the
lens resides in the bulge, then the mass range is much narrower:
$2.2\msun < M < 5.5\msun$ at $1\,\sigma$ and 
$1.34\msun < M < 8.3\msun$ at $2\,\sigma$. This suggests a low mass
black hole, although a neutron star or even a main sequence lens cannot
be ruled out.

\subsection{Constraint on a Possible Main Sequence Lens}
\label{sec-mslenses}

If the lens star is a main sequence star, then there is an additional
constraint on the distance and mass of the lens due to the observed
upper limit on the brightness of the lens star \citep{bennett-parbh}.
We have assigned a conservative upper limit on the V-band brightness of 
$V > 20.1\pm 0.5$ based upon the best parallax fit. We have applied this
constraint to the likelihood function by multiplying the 
likelihood function by the Gaussian probability that
the lens brightness exceeds upper limit on the lens star brightness.
The result is shown as the alternating short and 
long-dashed curve in Fig.~\ref{fig-masslike}, which indicates that
main sequence lens stars more massive than about $1.5\msun$ is ruled out.
A much tighter constraint on the brightness of the lens can be obtained
with HST images as has been done with the MACHO-96-BLG-5 black hole
candidate \citep{bennett-parbh}.

\section{Discussion and Conclusions}
\label{sec-con}

We have presented a microlensing parallax analysis of the 
MACHO-99-BLG-22/OGLE-1999-BUL-32 microlensing event using the publicly
available OGLE and MACHO data along with MPS and GMAN data.
The projected velocity indicated by the microlensing parallax fit 
has a direction that is nearly opposite of the direction of disk rotation.
This suggests that the lens is either very close to us in the Galactic
disk or in the Galactic bulge. The microlensing parallax mass-distance
relationship, \ref{eq-m}, indicates that the lens must be very
massive if it is nearby, with a mass of order $\sim 100\msun$. If the 
lens is in the Galactic bulge, however, its mass is much smaller,
$\sim 4\msun$. In both cases, the lens is likely to be a black hole.
These conclusions have been quantified with a likelihood analysis similar
to those presented by \citet{macho-par1} and \citet{bennett-parbh}.
We should note that this analysis does depend on the assumption that
stellar mass black holes are not born with a large velocity ``kick\rlap."
This assumption is supported by the available evidence \citep{bh-kick}, but
if this assumption is wrong, then the MACHO-99-BLG-22/OGLE-1999-BUL-32
could be a $\sim 10\msun$ black hole half way to the Galactic with a rotation
speed of only about $60\kms$.

We have not carried a Bayesian likelihood analysis with an assumed 
mass distribution prior as has been advocated by \citet{agol} and also
done by \citet{bennett-parbh}. Such an analysis is of little use for
this event because the lens mass is very likely to be $> 1.5\msun$.
Since a main sequence lens of this mass is excluded, we would have to
provide a prior distribution of stellar remnants, but 
there is virtually nothing known about mass function of stellar remnants
above $1.5\msun$.

Our analysis agrees with the previous analysis of \citet{ogle-99b22} 
and indicates that the MACHO-99-BLG-22/OGLE-1999-BUL-32 lens is very
likely to be a black hole, but we show that a nearby, massive black hole
of $\sim 100\msun$ and a low mass black hole of $\sim 4\msun$ in the
Galactic bulge are two distinct possibilities that can explain the
observed microlensing parallax parameters. When this event is
combined with the two other candidate black hole microlenses presented 
by the MACHO Collaboration \citep{bennett-parbh}, it appears that 
black holes may comprise a significant fraction of the mass of the Galactic
disk and bulge, perhaps as large as $\sim 10$\%. Future observations
with large ground based interferometers \citep{vlti} should be able to 
accurately determine the mass of future black hole microlenses and determine
the black hole mass fraction.

\acknowledgments

This work was supported, in part, 
through the NASA Origins Program Grant NAG5-4573.

\clearpage


\clearpage

%% If you are not including electronic art with your submission, you may
%% mark up your captions using the \figcaption command. See the 
%% No more than seven \figcaption commands are allowed per page, 
%% so if you have more than seven captions, insert a \clearpage 

%% Tables should be submitted one per page, so put a \clearpage before

%% Two options are available to the author for producing tables:  the
%% deluxetable environment provided by the AASTeX package or the LaTeX
%% table environment.  Use of deluxetable is preferred.
%%

%% Three table samples follow, two marked up in the deluxetable environment,
%% one marked up as a LaTeX table.

%% In this first example, note that the \tabletypesize{}
%% command has been used to reduce the font size of the table.
%% Note also that the \label command needs to be placed 

\clearpage

\begin{deluxetable}{lllllcccclrr}
\tabletypesize{\scriptsize}
\tablecaption{Microlensing Fit Parameters \label{tbl-fitpar}}
\tablewidth{0pt}
\tablehead{
\colhead{Fit} & \colhead{$f_{\rm OGLE}$} & \colhead{$f_{\rm MB}$} &
\colhead{$f_{\rm CTIO}$} & \colhead{$f_{\rm MPS}$} &  
\colhead{$t_0$ (MJD)} & \colhead{$u_0$} & 
\colhead{$\that$ (days)}   &  \colhead{$\vp$ (km/sec)} &
\colhead{$\theta$} & \colhead{${\chi^2\over {\rm(dof)}}$} & 
\colhead{$\chi^2$}
}
\startdata
Par. & $1.00(11)$ & $0.26(3)$ & $0.44(5)$ & $0.46(5)$ & $2745.0(1.5)$ & $0.043(6)$ & $1120(90)$ & $75(9)$ & $1.85(20)$ & $0.864$ & $899.8$ \\
Std. & $0.02(5)$ & $0.005(3)$ & $0.01(1)$ & $0.01(1)$ & $2742.4(1)$ & $0.0007(1)$ & $4.0(2)\times 10^4$ & $\equiv 0$ & $\equiv 0$ & $1.382$ & $1441.5$ \\
 \enddata

\end{deluxetable}

\begin{deluxetable}{lccccccc}
%% \tabletypesize{\scriptsize}
\tabletypesize{\small}
\tablecaption{Microlensing Parallax Likelihood Mass Estimates \label{tbl-mass}}
\tablewidth{0pt}
\tablehead{
 & \multispan{7}\hfil\  \ \
     {Confidence Levels $P(M/\msun < N)$ }\hfil  \\
\colhead{lens location} &
\colhead{$P=2.5\%$} & \colhead{$P=5\%$}  & \colhead{$P=16\%$} & 
\colhead{$P=50\%$}  & \colhead{$P=84\%$} & \colhead{$P=95\%$} & 
\colhead{$P=97.5\%$}
}
\startdata
disk or bulge & $560$ & $310$ & $106$ & $17.1$ & $3.2$  & $2.1$ & $1.70$ \\
disk          & $900$ & $480$ & $172$ & $49$   & $16.2$ & $8.5$ & $6.3$ \\
bulge         & $8.3$ & $7.3$ & $5.5$ & $3.6$  & $2.2$  & $1.60$ & $1.34$ \\
 \enddata

%% Text for table notes should follow after the \enddata but before
%% the \end{deluxetable}. Make sure there is at least one \tablenotemark
%% in the table for each \tablenotetext.

\end{deluxetable}

\end{document}